\documentclass[twocolumn,prb,showpacs,preprintnumbers,amsmath,amssymb]{revtex4}

\usepackage{subfigure}
\usepackage{graphicx}
\usepackage{dcolumn}
\usepackage{bm}
\usepackage{setspace}

\begin{document}


\title{Specific heat across the superconducting
dome in the cuprates
} 

\author{A. J. H. Borne$^{1,2,3}$}
\author{J. P. Carbotte$^{4,5}$}
\author{E. J. Nicol$^{1,2}$}%
\email{enicol@uoguelph.ca}
\affiliation{$^1$Department of Physics, University of Guelph,
Guelph, Ontario, Canada N1G 2W1} 
\affiliation{$^2$Guelph-Waterloo Physics Institute,
University of Guelph, Guelph, Ontario, Canada N1G 2W1}
\affiliation{$^3$PHELMA Grenoble INP, Minatec, 3 Parvis Louis N\'eel,
BP 275, 38016, Grenoble, Cedex 1, France}
\affiliation{$^4$Department of Physics and Astronomy, McMaster
University, Hamilton, Ontario, Canada L8S 4M1}
\affiliation{$^5$The Canadian
Institute for Advanced Research, Toronto, Ontario, Canada M5G 1Z8}

\date{\today}

\begin{abstract}
The specific heat of the superconducting cuprates is calculated
over the entire phase diagram. A $d$-wave BCS approach based on
the large Fermi surface of Fermi liquid and band structure theory
provides a good description of the overdoped region.
At underdoping it is essential to include  the emergence of a second
energy scale, the pseudogap and its associated Gutzwiller factor, which
 accounts for a reduction in the coherent piece of the
electronic Green's function due to increased correlations as
the Mott insulating state is approached. In agreement with experiment,
we find that the slope of the linear in $T$ dependence of the
low temperature specific heat rapidly increases above optimum doping while it
is nearly constant below optimum. Our theoretical calculations
also agree with recent data on Bi$_2$Sr$_{2-\rm x}$La$_{\rm x}$CuO$_{6+\delta}$
for which the normal state is accessed through the application of
a large magnetic field.
A quantum critical point is located at a doping slightly below optimum.

\end{abstract}


\maketitle

\section{Introduction}

The superconducting state of the underdoped cuprates shows anomalous
properties when compared with the case of optimum or overdoped.
There has been a considerable recent effort to understand these in a
model which includes the emergence of a pseudogap for doping $x$
below a quantum critical point (QCP) at $x=x_c$. The model of 
Yang, Rice, and Zhang\cite{yrz} (YRZ) on which this work is based,
has the pseudogap forming about the antiferromagnetic Brillouin zone
(AFBZ) boundary with its own characteristic energy scale. This model
is different in many respects from other competing order proposals such
as D-density waves\cite{ddw1,ddw2} and from a preformed pair model\cite{preform}
which involves a single energy scale. For a review of the successes
of the YRZ model in understanding the data, the reader is referred to
Schachinger and Carbotte\cite{ewaldyrz}. Many more details can be found in
references \cite{belen,jamesarpes,jamescv,kent,emilia,bascones,yrzarpes,yrzandreev,adriencv}
It is important to understand that this work goes beyond extensions of
BCS theory to include effects such as strong coupling
due to inelastic scattering\cite{mitrovic,daams,leung1,akis,hwang,tu,basov,nicolmfl,puchkov,nicol,ewald,inversion} and possible anisotropy\cite{tomlinson,leung2,donovan2,donovan3,branch,donovan1}
beyond a simple $d$-wave superconducting gap.

In a previous work\cite{jamescv}, we showed that the strong suppression
of the specific heat jump at $T_c$ and corresponding reduction in condensation
energy with increased underdoping can be understood as due to the emergence
of a pseudogap. Here we consider the equally anomalous observation that the 
slope of the linear in temperature law as $T\to 0$ is a strongly
increasing function of doping $x$ in the
overdoped regime while it is nearly constant at underdoping.\cite{loram,loram97,storey} While the overdoped
case is characteristic of a Fermi liquid, the underdoped case requires a new
ingredient for its understanding. In addition, we will also consider
the recent data of Wen et al.\cite{wen} on Bi$_2$Sr$_{\rm 2-x}$La$_{\rm x}$CuO$_{6+\delta}$
 for which $T_c$ at optimum is only 30K and thus superconductivity can
be quenched with a 9 Tesla magnetic field providing access to the
normal state as a function of doping in both Fermi liquid (overdoped)
and pseudogap (underdoped) state.

In Sec.~II, we present the essential elements of the theory of
YRZ\cite{yrz} needed for the calculation of the specific heat. Both
pseudogap state alone and with the addition of superconductivity within
a BCS  formulation are considered. For $x$ greater than $x_c$,
the doping at which the QCP associated with the pseudogap formation is set,
the Fermi surface is the usual large open contour of band theory and of
a Fermi liquid. For $x<x_c$, however, the Fermi surface reconstructs into
Luttinger pockets which progressively shrink in size as the Mott insulating
state is more closely approached. A consequence of this is that the density
of states at the Fermi level is reduced. In addition, a Gutzwiller
factor enters the theory which accounts for the depletion of the coherent
part of the electronic Green's function due to increased correlations which
eventually cause  the transition to an insulating state. In Sec.~III,
we present our results for the specific heat difference 
between superconducting and normal state as a function  of temperature
for values of doping ranging from 0.1 to 0.3 with $x=x_c=0.2$ the critical
doping at optimum which is also the QCP in our model.
The zero temperature limit of the specific heat difference is also considered
more explicitly and compared with the density of states
and the data of Wen et al.\cite{wen}. In Sec.~IV,
we show that while pseudogap formation strongly affects the overall
temperature dependence of the specific heat, it does not change its slope
at low $T$. This arises because this linear law only depends on the band structure
near the Dirac point on the heavily weighted part of the Fermi surface in the
nodal direction and this point is not changed by pseudogap formation in the
model of Ref.~\cite{yrz}. On the other  hand the size of the
slope depends directly on the Gutzwiller factor $g_t(x)$ which gives the
magnitude of the remaining coherent part of the Green's function.
It also depends on the ratio of the superconducting gap to critical
temperature which is expected to strongly increase with decreasing doping $x$
for $x<x_c$.\cite{ewaldyrz} These two effects combined lead to a rather
constant value of the slope over a significant range of doping. For overdoping,
Fermi liquid theory is recovered and in this case the slope shows much greater
variation
with $x$ as in experiment. Comparison with data is presented. In Sec.~V,
we provide a summary and give our conclusions.

\section{Formalism}

\begin{figure}
\includegraphics[width=0.48\textwidth]{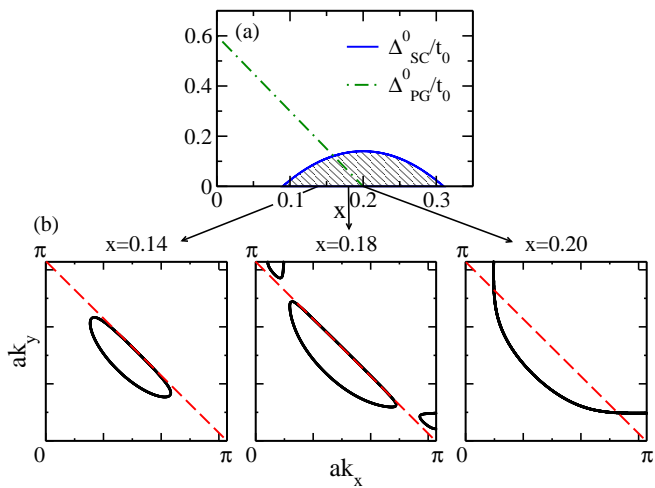}
\caption{\label{fig1} (Color online) 
(a) Phase diagram for the model discussed in the paper, showing the
pseudogap $\Delta^0_{\rm pg}$ and superconducting gap $\Delta^0_{\rm sc}$
in units of $t_0$ as a function of doping $x$. (b) Illustration of the
reconstruction of the Fermi surface in the first quadrant of the Brillouin
zone for doping values of $x=0.14$, 0.18 and 0.2. The red dashed line
is the AFBZ boundary.
}
\end{figure}

In the resonating valence bond spin liquid
model\cite{yrz}, the self-energy due to the pseudogap
is given by $\Delta_{\rm pg}^2(\boldsymbol{k})/[\xi^0(\boldsymbol{k})+\omega]$,
where $\xi_{\boldsymbol{k}}^0  =  - 2t(x)(\cos k_xa  +
\cos k_ya)$ with $t(x)$ the effective first neighbor hopping parameter
and $\Delta_{\rm pg}(\boldsymbol{k})$ is a pseudogap which has $d$-wave
symmetry in the Brillouin zone. It is given by 
\begin{equation}
\Delta_{\rm pg}(\boldsymbol{k})=\frac{\Delta_{\rm pg}^{0}(x)}{2}(\cos
k_xa -\cos k_ya).
\label{eq:gapk}
\end{equation}
Here the amplitude $\Delta^0_{\rm pg}(x)$ is linear in doping $x$
as shown in Fig.~\ref{fig1}(a) 
where the superconducting
dome is also shown for easy orientation. In the above, $\boldsymbol{k}$
is the momentum and  $a$ is the lattice constant of the CuO$_2$ plane. Without
superconductivity, the coherent part of the electronic Green's function is given by
\begin{equation}
G(\boldsymbol{k},\omega)=\frac{g_t(x)}{\omega-\xi(\boldsymbol{k})-
\Delta_{\rm pg}^2(\boldsymbol{k})/[\xi^0(\boldsymbol{k})+\omega]},
\end{equation}
where $\xi(\boldsymbol{k})$ is the electron dispersion curve of band theory
and $g_t(x)$ is a Gutzwiller factor equal to $2x/(1+x)$. This latter quantity
 accounts
for the effect of correlations which reduces the weight of the coherent
part of $G(\boldsymbol{k},\omega)$ and adds an incoherent background not 
considered in this work. It
 also enters the renormalized band structure
dispersion curve which includes up to third nearest neighbor hopping
 and narrows
as  the Mott transition is approached with decreasing value of
$x$. This narrowing is modeled by a second Gutzwiller factor
$g_s(x)$ in addition to $g_t(x)$. 
 Effectively, for a given $\boldsymbol{k}$,
 there are two electron branches $E^\pm_{\boldsymbol{k}}$ 
with weights $W^\pm_{\boldsymbol{k}}$ given by
\begin{equation}
E^\pm_{\boldsymbol{k}}=\frac{\xi_{\boldsymbol{k}}-\xi^0_{\boldsymbol{k}}}{2}
\pm\sqrt{\biggl(\frac{\xi_{\boldsymbol{k}}+\xi^0_{\boldsymbol{k}}}{2}\biggr)^2+
\Delta^2_{\rm pg}(\boldsymbol{k})}
\end{equation}
and 
\begin{equation}
W^\pm_{\boldsymbol{k}}=\frac{1}{2}\biggl[1\pm
\frac{(\xi_{\boldsymbol{k}}+\xi^0_{\boldsymbol{k}})/2}
{\sqrt{[(\xi_{\boldsymbol{k}}+\xi^0_{\boldsymbol{k}})/2]^2+
\Delta^2_{\rm pg}(\boldsymbol{k})}}\biggr].
\end{equation}
In terms of  $E^\pm_{\boldsymbol{k}}$, 
the Fermi surface contours of 
zero excitation energy are given by $E^\pm_{\boldsymbol{k}}=0$ and
these are shown in Fig.~\ref{fig1}(b) for three values of
doping $x=0.14, 0.18,$ and 0.2. In the first two, there is a hole pocket
centered about the nodal direction $\theta=\pi/4$. This pocket is
determined from the equation $E^-_{\boldsymbol{k}}=0$ and 
$E^-_{\boldsymbol{k}}$ is positive only for momenta falling within the
area defined by the hole pocket. The size of the pocket 
shrinks as $x$ decreases and
we come closer to half filling and the Mott insulating state. For the
case $x=0.18$, close to the QCP at $x=0.2$
where pseudogap formation starts 
in our model, there is an additional electron pocket near the corner
of the AFBZ. This pocket
is determined by the equation $E^+_{\boldsymbol{k}}=0$. Both electron and
hole pockets have two sides, one weighted by 
$W^\pm_{\boldsymbol{k}}$  of order one and the other, which takes on a 
close resemblance to the AFBZ, has only a small weight in comparison.
This small weight goes to zero in the limit of no pseudogap and the energy
$E^\pm_{\boldsymbol{k}}$ becomes the Umklapp energy surface 
$\xi^0_{\boldsymbol{k}}$.
On the other hand, in this same limit, the heavily weighted part
traces out the large Fermi surface of Fermi liquid theory (Fig.~\ref{fig1}(b),
far right panel) which has weight
one everywhere. These remarks make clear the evolution from large Fermi
surface into small hole pockets. As the pockets shrink in size, the number of states
which carry excitations of zero energy becomes small. Also, it should be kept
in mind that because of the small weight on the backside of the hole
pocket, we are effectively dealing with an arc when considering many
properties.

When superconductivity is included in a BCS formalism, the electronic
spectral density $A(\boldsymbol{k},\omega)$ can be written in the
form
\begin{equation}
A(\boldsymbol{k},\omega)=\sum_{\alpha=\pm}g_t(x)W_{\boldsymbol{k}}^\alpha
[(u_{\boldsymbol{k}}^\alpha)^2\delta(\omega-E^\alpha_{\boldsymbol{k},S})+(v_{\boldsymbol{k}}^\alpha)^2\delta(\omega+E^\alpha_{\boldsymbol{k},S})],
\label{eq:A}
\end{equation}
with  $E^{\alpha}_{\boldsymbol{k},S}=\sqrt{(E_{\boldsymbol{k}}^{ \alpha })^2  +
  \Delta _{\rm sc}^2(\boldsymbol{k})}$. Here, the superconducting gap
  $\Delta _{\rm sc}(\boldsymbol{k})$ 
is assumed to have the same $d$-wave dependence
in momentum space as in Eq.~(\ref{eq:gapk}) for the pseudogap with amplitude
$\Delta_{\rm sc}^0(x)$ replacing $\Delta_{\rm pg}^0(x)$ and
$\Delta_{\rm sc}^0(x)$ is assumed to have the same doping dependence
as the critical temperature $T_c(x)$ dome. This is shown  in 
Fig.~\ref{fig1}(a) and for definiteness
we will assume in all our numerical
work that $2\Delta^0_{\rm sc}(x)/k_BT_c(x)=6$. When later
we compare with experimental data, we will remove this simplifying assumption.
The Bogoliubov weights in Eq.~(\ref{eq:A}) are 
\begin{eqnarray}
(u^\alpha_{\boldsymbol{k}})^2&=&\frac{1}{2}\biggl(1+\frac{E_{\boldsymbol{k}}^\alpha}{E^\alpha_S}\biggr),\\
(v^\alpha_{\boldsymbol{k}})^2&=&\frac{1}{2}\biggl(1-\frac{E_{\boldsymbol{k}}^\alpha}{E^\alpha_S}\biggr).
\end{eqnarray}
The dispersion curves for $\xi_{\boldsymbol{k}}$ and $\xi^0_{\boldsymbol{k}}$
are taken from the work of Ref.~\cite{yrz} and are unchanged here as are the other 
parameters, namely,
\begin{equation}
\Delta^{0}_{\rm
    sc}(x)=0.14t_0[1-82.6(x-0.2)^2]
\label{eq:delscx}
\end{equation}
and 
\begin{equation}
\Delta^{0}_{\rm
    pg}(x)=3t_0(0.2-x), 
\end{equation}
with $t_0$ an unrenormalized nearest neighbor hopping parameter
characteristic of the CuO$_2$ plane. 

There are two equivalent ways for calculating the specific heat
$C_V(T)\equiv\gamma(T)T$. 
One is through the entropy $S(T)$,
the other through the internal energy $U(T)$. The entropy is
\begin{eqnarray}
S(T)=-2k_Bg_t(x)\sum_{\boldsymbol{k},\alpha=\pm}&&W_{\boldsymbol{k}}^\alpha\{f(E^\alpha_{\boldsymbol{k}S})\ln f(E^\alpha_{\boldsymbol{k}S})\nonumber\\
&&+[1-f(E^\alpha_{\boldsymbol{k}S})]\ln[1-f(E^\alpha_{\boldsymbol{k}S})]\},\nonumber\\
&&
\label{eq:S}
\end{eqnarray}
where $f(x)$ is the Fermi-Dirac temperature distribution function. Alternatively,
the internal energy can be expressed in terms of the single spin
 density of states
$N(\omega)$ as 
\begin{equation}
U(T)=2\int_{-\infty}^\infty \omega N(\omega)f(\omega)d\omega
\label{eq:U}
\end{equation}
where the two is for spin degeneracy and the density of
states is given by
\begin{equation}
N(\omega)=\sum_{\boldsymbol{k},\alpha=\pm}g_t(x)W_{\boldsymbol{k}}^\alpha
[(u_{\boldsymbol{k}}^\alpha)^2\delta(\omega-E^\alpha_{\boldsymbol{k},S}
)+(v_{\boldsymbol{k}}^\alpha)^2\delta(\omega+E^\alpha_{\boldsymbol{k},S})].
\label{eq:DOS}
\end{equation}
In both Eqs.~(\ref{eq:S}) and (\ref{eq:DOS}), the sum is
over the entire Brillouin zone and 
$C_V(T)=dU(T)/dT=\gamma(T)T$.

\section{Numerical Results for the Specific Heat}

\begin{figure}
\includegraphics[width=0.48\textwidth]{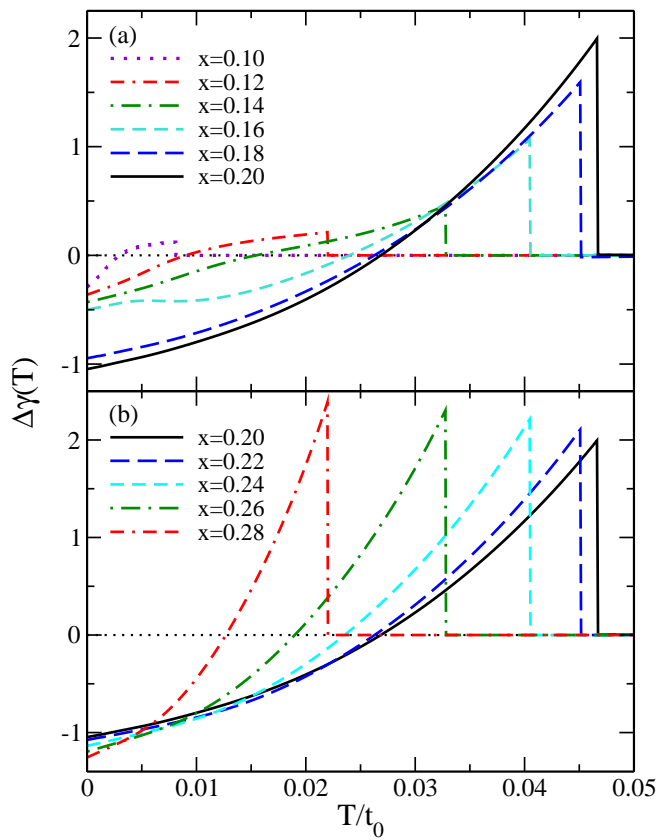}
\caption{\label{fig2} (Color online) 
$\Delta\gamma(T)$ versus $T/t_0$, where 
$\Delta\gamma\equiv\gamma_{sc}-\gamma_n$
is in units of $k_B^2/t_0$ per volume. Curves are shown for (a) underdoped
and (b) overdoped cases along with optimal doping for reference.
}
\end{figure}

In Fig.~\ref{fig2}, we present our results for the difference between
the superconducting and normal state specific  heat $\gamma(T)$,
namely $\Delta\gamma(x,T)\equiv\gamma_{sc}(x,T)-\gamma_n(x,T)$
in units of $k_B^2/t_0$ per volume as a function
of temperature $T$. Several values of doping $x$ are considered
as indicated in the figure. The top frame covers the underdoped regime,
while the bottom frame is for overdoped. As the superconducting dome
given in Eq.~(\ref{eq:delscx}) is symmetric in doping about optimum
$x=0.2$, the top and bottom frame curves come in pairs with the same
value of critical temperature (with the exception of
not displaying $x=0.3$ for clarity). 
Comparison of top and bottom curves in a 
given pair shows that the formation of a pseudogap and associated Fermi
surface reconstruction [Fig.~\ref{fig1}(b)] provides a drastic suppression
of the jump at the critical temperature $T_c$ and also of the slope
just below $T_c$. This is in qualitative agreement with 
experiment\cite{wen,loram,loram97,storey}, as discussed already in 
Ref.~\cite{jamescv}. These results are clearly not part of ordinary
$d$-wave BCS theory where instead the jump is large and relatively independent
of doping as seen in the Fig.~\ref{fig2}(b). For reference in assessing
these curves, we recall that for a constant density of states model with
a superconducting $d$-wave
gap defined on the Fermi surface, the canonical value of
the jump $\Delta\gamma(T_c)/\gamma_n(T_c)$ is 0.95 for a gap to critical
ratio $2\Delta^0_{\rm sc}/k_BT_c=4.3$ and is the same for all superconductors
regardless of the size of $T_c$. Here it deviates from this universal
law because our superconducting gap is defined in the entire Brillouin
zone according to Eq.~(\ref{eq:gapk}) with the pseudogap replaced by 
$\Delta^0_{\rm sc}$ and our energy bands can be complicated even when
the large Fermi surface of Fermi liquid theory is involved. Also, the magnitude
of the jump itself is increased because we have used a gap to $T_c$
ratio of 6 rather than the weak coupling limit of 4.3 and this has resulted 
in a change of $\Delta\gamma(T_c)/\gamma_n(T_c)$ from 0.95 to $\sim 1.7$.
 Thus, the
large decrease in the jump $\Delta\gamma(T_c)$ seen in Fig.~\ref{fig2}(a) is a 
direct consequence of pseudogap formation and accompanying Fermi surface
reconstruction.

\begin{figure}
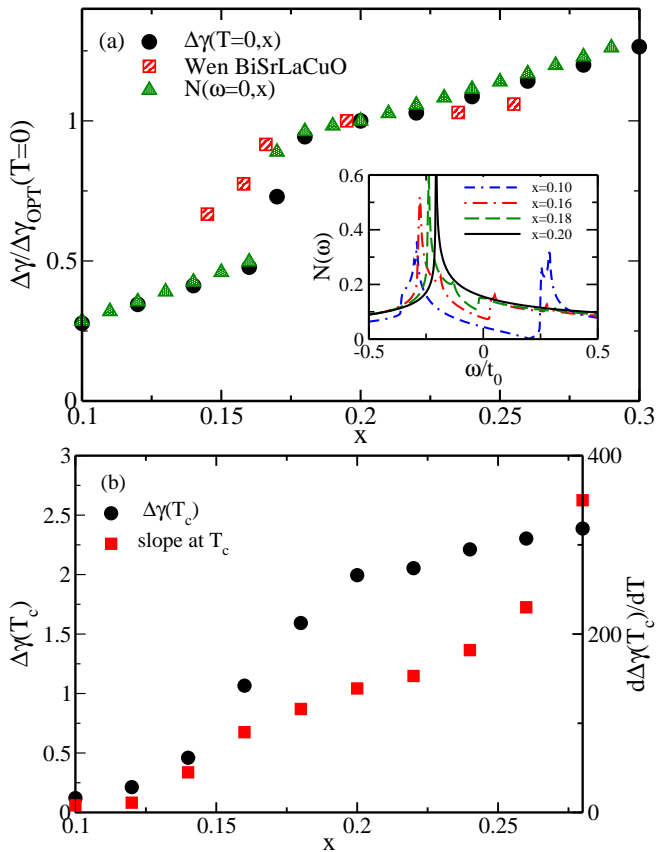

\includegraphics[width=0.48\textwidth]{fig3a.eps}
\includegraphics[width=0.48\textwidth]{fig3b.eps}
\caption{\label{fig3} (Color online) 
(a) $\Delta\gamma$ as $T\to 0$ normalized to the same quantity
at optimal doping versus doping $x$. Values extracted from the data
of Wen et al.\cite{wen} are shown along with the DOS at zero energy $N(0)$
similarly normalized.
The inset shows $N(\omega)$ in units of 
$t_0$  for several dopings
as a function of $\omega/t_0$. $N(\omega)$ and $N(0)$
refer to the normal state with a pseudogap. 
(b) Value of $\Delta\gamma$ at $T_c$
 from Fig.~\ref{fig2} (left axis) and slope just below $T_c$,
$d\Delta\gamma(T_c)/dT$ (right axis) as a function of $x$.
}
\end{figure}

The size of the specific heat difference seen at zero temperature
is also of interest and is very different in the underdoped than in
the overdoped regime. While for near optimum, optimum and 
overdoped 
cases the value of $\Delta\gamma(T)$ as $T\to 0$ is around -1 in our units,
 for the underdoped cases, it has moved instead to a value of roughly
$\sim -0.5$. In this limit $\Delta\gamma(T=0)$ 
simply reduces to its normal state value $\gamma_n$ and 
is a direct measure of the
value of the density of states $N(\omega)$ at the Fermi energy, {\it i.e.}
 at $\omega=0$. In Fig.~\ref{fig3}(a), we compare the results 
of our calculations 
for $\Delta\gamma(T=0)/\Delta\gamma_{\rm OPT}(T=0)$
(circles) with the specific heat data of Wen et al.\cite{wen} on
Bi$_2$Sr$_{2-\rm x}$La$_{\rm x}$CuO$_{6+\delta}$
(squares). 
Here, OPT refers to the value  taken at optimal doping ($x=0.2$ in our model). 
In carrying out this comparison, we have
fit a parabolic form to the data of Wen et al.  for $T_c$ versus doping.
We found that the fit which best captured the data
was when optimum was taken at their doping of  $p=0.165$
and $t_0$ is set to 56 meV for $2\Delta_0/k_BT_c=6$. 
Similar, but less ideal 
fits would also work for optimum at $p=0.16$. Note that this material
does not have a very high $T_c$ and consequently has a narrower dome
compared with what is typically found in the literature. It was chosen
for experiment because reasonable magnetic fields can be used to suppress
$T_c$ to zero and so access the normal state. Given our fit, we have
shifted the value of $p$ in experiment by 0.035
to get correspondence with our theoretical work for which $x$ at 
optimum is kept at $x=0.2$. Further discussion of this is presented
in Sec.~IV.  We have kept the pseudogap line to be as shown 
in Fig.~\ref{fig1}(a)
which coincides with optimum doping, but this could easily
be changed. Returning to 
Fig.~\ref{fig3}(a), both theory and experiment show two distinct regimes:
a somewhat flat or slowly rising region 
above $x\simeq 0.17$ with a sharp drop 
below this value. The behavior in  the region above
$x=0.2$ is sensitive to the choice of bandstructure parameters as shown in 
the original work of  YRZ (Ref.\cite{yrz}).  Because the data
is fairly flat in this region, we have opted to alter the bandstructure
accordingly and have allowed the $t'$ and $t''$ parameters to
continue to vary with $x$ rather than to become constant for
$x>0.2$ as in Ref.~\cite{yrz}. Returning to the rapid drop around
$x\sim 0.17$, we find
that it is displaced
to slightly lower values of $x$ in the data as compared to theory
and we take this to mean that in 
Bi$_2$Sr$_{2-\rm x}$La$_{\rm x}$CuO$_{6+\delta}$
the QCP associated with the start of pseudogap formation occurs at
a doping level close but slightly below optimum. It is important
to realize that in the present model, the point at which the value of
the density of states at the Fermi level $N(0)$ begins to 
be strongly reduced by the growth of the pseudogap, is not exactly at the
QCP but is instead displaced to smaller $x$ values. The physics of this
displacement is easily understood with reference to the inset in
Fig.~\ref{fig3}(a).
First we show, in the main frame, that
 our numerical results for $N(0)$ versus $x$ (triangles)
 are very close to the solid black circles as they must be. These were
obtained not from an extrapolation to zero temperature
of a specific heat calculation as were the solid circles but 
instead from the full DOS
$N(\omega)$ versus $\omega$ shown in the inset.
This good agreement
is  taken as providing a check on our numerical
work. More significantly 
it is important to realize that the Fermi surface reconstruction
from the large open Fermi liquid surface of Fig.~\ref{fig1} ($x=0.2$) 
to a Luttinger hole pocket does not immediately lead to change in the
value of $N(0)$. In fact, as can be seen clearly in the inset
of Fig.~\ref{fig3}(a), at $x=0.18$ the DOS at $\omega=0$
is hardly changed from its Fermi liquid value (see solid black
curve for $x=0.2$ for comparison). 
Rather the effect of the pseudogap is to provide 
 a depletion of states at negative $\omega$. For
$x=0.16$ however, the upper edge of this depletion region which is rather
sharp, has moved across the Fermi energy and  $N(0)$ has
become significantly reduced as shown in the triangles
of the main frame in Fig.~\ref{fig3}(a). This is not
surprising. The DOS at $\omega=0$ depends on the Luttinger contours
of zero energy shown in Fig.~\ref{fig1}. At $x=0.18$, there are both
hole and electron pockets and the number of states having zero energy
is not very different from the number when the large open Fermi surface
applies. It is only when we reach a single small hole pocket and no
electron pockets that depletion of states at $\omega=0$ becomes
significant. On the face of it, one might even think that more zero energy
states are involved when  $x=0.18$ than when $x=0.2$ because the length
of the Fermi surface is larger when there are pockets. But this is not
so because, as we have already stated, parts of the Luttinger contours
carry very little weight. 

In Fig.~\ref{fig3}(b), we show results for the 
jump at $T_c$ of the specific heat $\Delta\gamma(T_c)$
(circles)
and its slope out of $T_c$ (squares)
 as a function of doping.
We note that for the jump
$\Delta\gamma(T_c)$,
 a notable drop occurs almost immediately below optimum $x=0.2$,
which can be used to identify the QCP. For the slope,
the signature of the QCP is not as sharp, only a change in
curvature arises at $x=0.2$.
These results are to be contrasted with those in Fig.~\ref{fig3}(a).
If one were to estimate the value of doping to be associated with
the QCP from the behavior of $\Delta\gamma(T)$ at $T\to 0$, it is necessary
to account for a significant displacement downward towards smaller values
of $x$ of the rapid drop in $N(0)$ as compared to the value of $x$
at the QCP. Thus
extracting a QCP
from thermodynamics requires some care but it clearly can
be done either from the jump at $T_c$ or the value of the difference 
$\Delta\gamma(T)$ as $T\to 0$. 

\section{Low temperature behavior}

\begin{figure}
\includegraphics[width=0.48\textwidth]{fig4atod.eps}
\includegraphics[width=0.48\textwidth]{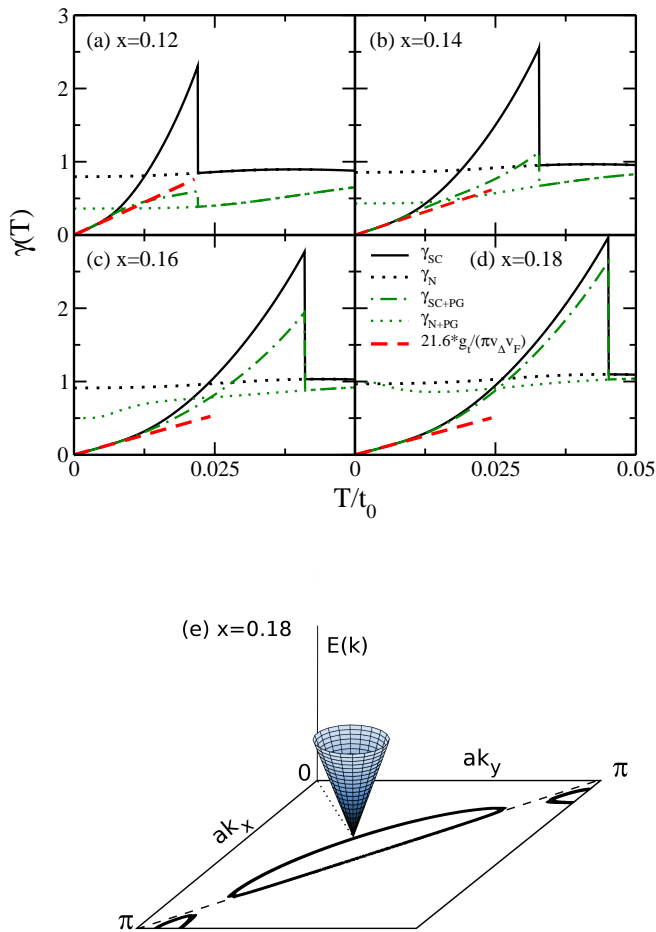}
\caption{\label{fig4} (Color online) 
$\gamma(T)$ versus $T/t_0$ 
shown in the absence of any gaps $\gamma_N$,
with
superconductivity only $\gamma_{SC}$, pseudogap only $\gamma_{PG}$ and
with both gaps present $\gamma_{SC+PG}$. The red dashed line gives
the theoretical expression for the slope from Eq.~(\ref{eq:slope}).
(a)-(d) show a range of dopings for the underdoped case.
(e) Schematic of the superconducting Dirac cone shown on the 
side of the $x=0.18$
Fermi pocket with large quasiparticle weight.
}
\end{figure}

We now consider the low temperature behavior of the superconducting
state specific heat $\gamma_{sc}(T)$ which  is
emphasized in Fig.~\ref{fig4}(a)-(d)
for underdoped cases.
 We base our discussion on the physics of the schematic shown in
Fig.~\ref{fig4}(e).
 The Luttinger pockets
are shown as heavy black lines for $x=0.18$
which has electron as well as hole areas. Also shown is a Dirac cone
centered on the highly weighted Fermi surface contour in the nodal
direction. The cone illustrates the quasiparticle energies at low energy
in the
superconducting state as a function of $k_x$ and $k_y$ in the
Brillouin zone. Only the upper right quadrant is depicted. At very low temperatures,
the tip of this cone is the only region in momentum space
where there is a finite thermal occupation of excited quasiparticles.
Thus, the specific heat can depend only on characteristic parameters
associated with the Dirac point. But this point is particularly
simple. It corresponds to $E^-_{\boldsymbol{k}}=0$ in the
normal pseudogap case and does not change with the onset of superconductivity.
In the nodal direction, the pseudogap $\Delta_{\rm pg}({\boldsymbol{k}})=0$
and therefore $\xi_{\boldsymbol{k}}=0$ which is the condition for the
underlying large Fermi surface. Note that strictly speaking
$\xi_{\boldsymbol{k}}$ involves the chemical potential $\mu_p$ associated
with the case when the pseudogap is present and is slightly different
from the chemical potential of Fermi liquid theory. In the former case,
it is determined from the Luttinger sum rule while in the latter
case of a large Fermi surface one might determine $\mu_p$ from
the DOS filling up to ${\boldsymbol{k}}={\boldsymbol{k}}_F$. Neglecting
this small difference, the band energy at the Dirac point is unchanged from
its Fermi liquid value and the well known techniques\cite{durst} for obtaining
the $\omega\to 0$ limit of the DOS $N(\omega)$ apply unaltered. The
result is
\begin{equation}
N(\omega)\simeq|\omega|\frac{g_t(x)}{\pi v_Fv_\Delta},
\label{eq:Nlow}
\end{equation}
where $v_F$ and $v_\Delta$ are the Fermi and gap velocity, respectively.
Note that Eq.~(\ref{eq:Nlow}) predicts that the only effect of the
pseudogap formation on the DOS around the Dirac point is the appearance
of the Gutzwiller factor $g_t(x)$. While to a good approximation $v_F$
is unchanged, the gap velocity can be changed in magnitude if the
ratio of zero temperature gap to critical temperature $T_c$ is affected
by pseudogap formation as was found in the recent work of 
Schachinger and Carbotte\cite{ewaldyrz}. 
This provides a second important change in Eq.~(\ref{eq:Nlow}) as
compared with the more familiar Fermi liquid case.

Inserting Eq.~(\ref{eq:Nlow}) into Eq.~(\ref{eq:U}) for the internal
energy $U(T)$, we obtain the simple analytic result for $\gamma_{sc}(T)$
in the limit of low temperature:
\begin{equation}
\gamma_{sc}(T)=4k_B^2 (k_BT)\frac{g_t(x)}{\pi v_Fv_\Delta} h,
\label{eq:slope}
\end{equation}
where $k_B$ is the Boltzmann constant and $h$ is a number
given by
\begin{equation}
h=4\int^{+\infty}_{-\infty} dy|y|^3\cosh^{-2}(y)\simeq 5.4.
\end{equation}
We have obtained, as in ordinary $d$-wave BCS theory, a linear
in $T$ law with the same material factors appearing, {\it i.e.} $v_F$ and 
$v_\Delta$, but with an extra Gutzwiller factor of $g_t(x)$. We also need to
note that $\Delta$ can be affected by the variation with doping $x$ of the
gap ratio $R(x)\equiv 2\Delta^0_{\rm sc}(x)/k_BT_c(x)$.

In Fig.~\ref{fig4}, we show our numerical results for $\gamma_{sc}(T)$
at four values of $x$ namely 0.12, 0.14, 0.16 and 0.18. In all cases,
the solid line is for the case when no pseudogap is included while the
dashed-dotted includes a pseudogap with its magnitude chosen to
correspond to the value of doping $x$ chosen. Otherwise, there are
no other changes. Gutzwiller factors are included in these curves
and the ratio $R(x)$ is fixed at value 6 as in all other
numerical work presented in this paper. While the inclusion of a pseudogap
has drastic effects on $\gamma_{sc}(T)$, the slope as $T\to 0$ is
completely unaffected  and this slope agrees perfectly
with the simplified analytic results of Eq.~(\ref{eq:slope})
given as the red dashed line. This constitutes an important 
prediction of YRZ theory and will be verified later when we make
comparison with experimental data.

The heavy dotted lines in Fig.~\ref{fig4} show the normal state specific
heat $\gamma_N(T)$ in the Fermi liquid
and the light dotted, with a pseudogap included
$\gamma_{N+PG}(T)$. It is instructive to consider these in some detail
and in particular to describe how they are related to the detailed
variation of $N(\omega)$ versus $\omega$ in the pure pseudogap
state, {\it i.e.} with no superconductivity. We begin with the case
$x=0.18$ for which the pseudogap is small and the Fermi contours include 
electron as well as hole pockets. In this case, the light dotted curve
falls below the heavy dotted one but rises to meet it as $T\to 0$.
This behavior can be traced to the $\omega$ variation seen in the inset
of Fig.~\ref{fig3}(a) which shows $N(\omega)$ versus $\omega$.
The solid black curve forms a reference and is the Fermi liquid result
for $x=0.2$, zero pseudogap. Comparing with the long-dashed
green curve for $x=0.18$, we note no visible change in the value
of $N(\omega=0)$ but there is a significant
depression of DOS at negative energies. As $T\to 0$ only $N(0)$ is sampled in
the specific heat and hence both cases, with and without a pseudogap, agree.
At small but finite $T$, however, the light-dotted curve falls below
the heavy dotted line in Fig.~\ref{fig4}(d) because the
dip in its DOS at $\omega<0$ starts to be sampled and this reduces the 
specific heat. Specific heat
 is, however, a rather broad spectroscopy for $N(\omega)$
because the thermal factor in $\gamma(T)$ in Eq.~(\ref{eq:U}) samples of order 
$5k_BT$ or so about $\omega=0$. The case $x=0.12$ is also noteworthy. In this
instance the DOS about the Fermi energy is nearly monotonic although
depressed in value as compared to $x=0.2$ by about 50\%
and this correspondingly reduces the value of $\gamma_N(T)$
by the same amount with little other changes.

\begin{figure}
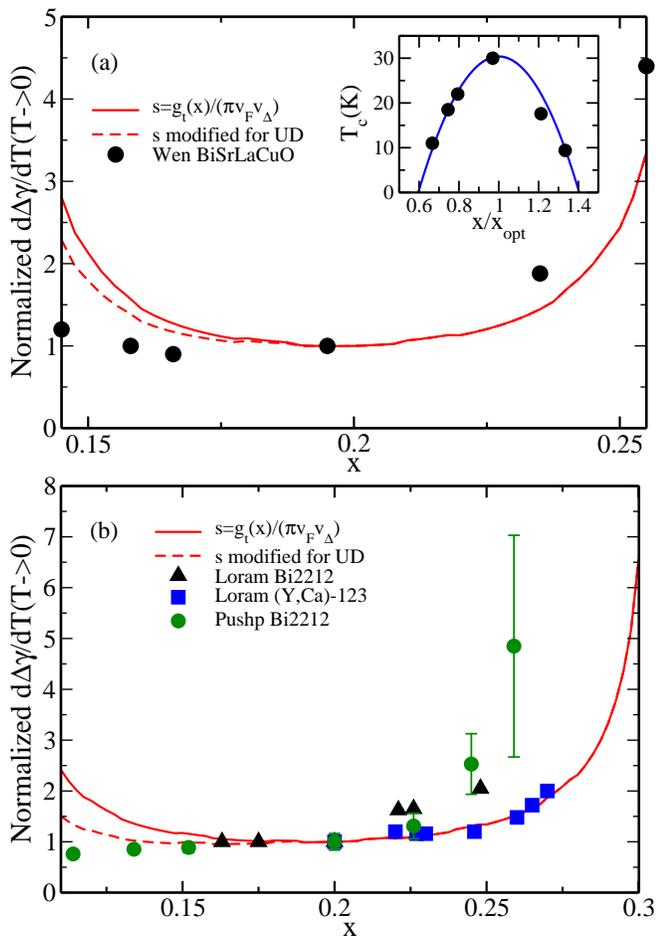

\includegraphics[width=0.48\textwidth]{fig5a.eps}
\includegraphics[width=0.48\textwidth]{fig5b.eps}
\caption{\label{fig5} (Color online) 
The slope of the specific heat at low $T$ normalized to the value 
at optimal doping. Frame (a) shows the data extracted from Wen et al.\cite{wen}
compared with the theory of Eq.~(\ref{eq:slope}) for
the $T_c$ dome given in the inset where we have fit a functional form
for  $x$
to their $T_c$ data  (see text for discussion). The dashed curve includes the
correction $R(x)$ discussed in the text. (b) The extracted data of 
Loram et al.\cite{loram,loram97,storey} is compared to the theoretical slope.
Shown also is the slope data from the STS DOS of Pushp et al.\cite{pushp}. 
}
\end{figure}

We turn next to a  comparison with data on the slope of the specific heat in
the $T\to 0$ limit. These are presented in Fig.~\ref{fig5}(a) 
for the Bi$_2$Sr$_{2-\rm x}$La$_{\rm x}$CuO$_{6+\delta}$ data of
Wen et al.\cite{wen} and in (b) for the data on Bi-2212 
and (Y,Ca)-123 of Loram et al.\cite{loram,loram97,storey} where we
have read slopes from the published figures as best as was possible. A more
thorough analysis by the experimentalists might achieve better accuracy. In the
inset of Fig.~\ref{fig5}(a), we show our results for the fit to the
$T_c$ dome of the Wen et al.\cite{wen} data. We find 
\begin{equation}
k_BT_c=\frac{0.14}{3}[1-225(p-0.165)^2]t_0,
\end{equation}
which provides a good representation of the $T_c$ data as a function
of doping $p$. We then take $x=p+0.035$ to place the data on our curves.
 In the main frame, the solid black circles are the
data for the slope at $T\to 0$ normalized to its value at optimum doping.
The solid red line gives results of Eq.~(\ref{eq:slope}) with the
assumption of $R=6$. The agreement is very good and the sharp rise
in slope in the overdoped region is captured by our model. For the
deeply underdoped case, our theoretical values are somewhat higher than
experiment, but as we have already mentioned, the
gap ratio may well vary with doping. Solving a BCS-like pairing
equation with pseudogap formation and corresponding Fermi surface reconstruction
accounted for, Schachinger and Carbotte\cite{ewaldyrz} found that (for $x\leq 0.2$) approximately
\begin{equation}
R(x)\sim  4.3[1+75(x-0.2)^2].
\end{equation}
If this correction, scaled to 6 at optimum,
is incorporated into the comparison with experiment,
we get the red dashed curve which agrees better with the data at small $x$.
Note finally that to compare data with theory, we have shifted all $x$ values
in Fig.~\ref{fig5}(a) by 0.035 because we wished to remain, as in the paper
of YRZ\cite{yrz}, with optimum doping at $x=0.2$ rather than the
experimental value of about 0.165. This is also true for Fig.~\ref{fig5}(b)
where we compare with data on Bi2212 (triangles) and (Y,Ca)-123
(squares). The agreement with theory (solid and dashed red curves)
is again good. We have also included one further comparison with 
scanning tunneling spectroscopy (STS) results on Bi2212 by Pushp et al.\cite{pushp}
(circles). STS gives the DOS and not the specific heat, but this latter
quantity follows directly from a knowledge of $N(\omega)$. We can
use the STS data in the limit of $\omega\to 0$ to determine $N(\omega)$
as in Eq.~(\ref{eq:Nlow}) and so get
$\gamma_{sc}(T)$ for $T\to 0$. These are the solid circles which
provide a confirmation of the specific heat results and
also provide a significant cross check between these two important
but very different probes of the microscopic structure of the 
superconducting state in the underdoped cuprates and indeed over the
entire phase diagram.

\section{Summary and Conclusions}

The behavior of the specific heat of the underdoped cuprates differs
profoundly from that observed on the overdoped side of their phase
diagram. At optimum and overdoping, an ordinary BCS approach based on a
Fermi liquid normal state with constant DOS provides a first reasonable
understanding on the assumption that the gap has $d$-wave symmetry.
This ensures a linear in $T$ low temperature law for the specific heat
$\gamma(T)$ with $C_V(T)=\gamma(T)T$. It also gives a jump at $T_c$
normalized  to its normal state value of 0.95 and this value can
be increased if the gap to $T_c$ ratio is changed from 4.3 to a higher
value. On the deeply underdoped side, however, there is new physics
which cannot be described even when strong coupling effects\cite{mitrovic,daams,leung1,akis,hwang,tu,basov,nicolmfl,puchkov,nicol,ewald,inversion} 
due to inelastic 
scattering\cite{hwang,tu,basov,nicolmfl,puchkov,nicol,ewald,inversion} 
 are accounted for  and/or effects
of anisotropy\cite{tomlinson,leung2,donovan2,donovan3,branch,donovan1}.
which go beyond a simple lowest harmonic $d$-wave picture for the 
superconducting gap.

In the resonating valence bond spin liquid picture, a second energy
scale, the pseudogap, emerges and grows in magnitude as $x$ is 
reduced towards the Mott insulating state. This pseudogap leads to a
loss of metallicity. It also radically reduces
the size of the specific heat jump
at $T_c$ as noted in experiment\cite{loram,loram97,storey} and also
in theory\cite{jamescv}. However, as we show here the limit of $T\to 0$
of $\gamma_{sc}(T)$ is not directly affected by the size of 
$\Delta^0_{\rm pg}$ because this limiting value depends only on the band
structure and superconducting gap right at the Dirac point in the 
Brillouin zone. But this point is not importantly changed by pseudogap
formation and Fermi surface
reconstruction so
 that the formula for the slope $d\gamma_{sc}(T)/dT$ as $T\to 0$
remains unchanged in form but with  two important modifications. 
First, in the
resonating valence bond spin liquid, there appears a Gutzwiller factor
$g_t(x)$ which depends strongly on doping $x$ and represents the remaining
weight in the
coherent part of the electronic Green's function as correlations become
more important and consequently shift more spectral weight into an incoherent
background at higher energies. 
  A second factor is that the ratio of 
$2\Delta^0_{\rm sc}(x)/k_BT_c(x)=R(x)$ can vary with doping. For overdoped and
optimally doped, in ordinary BCS $d$-wave it has a value of 4.3, but
Schachinger and Carbotte\cite{ewaldyrz} have found that it rises
considerably in the underdoped region of the phase diagram. These
authors solve a BCS gap equation generalized  to include pseudogap
formation and attendant Fermi surface reconstruction. The results of
such a theoretical study show that $R(x)$ increases beyond a value of 7
before  the lower end of the superconducting dome is reached. These two
effects, along with the linear in $T$ law which we have shown to still hold
in YRZ theory at underdoping, allow us to understand a
previously anomalous feature of
the data namely the slope remains reasonably constant at underdoping
while it increases sharply in the overdoped region.
This last observation is consistent with a slope
which varies inversely as the gap and this gap
decreases toward zero as we approach 
the upper end of the dome. Comparison between theory and
data on BiSrLaCuO, Bi2212, and (Y,Ca)-123 show good agreement. A further comparison
was made with STS data which provides information on the average density of
quasiparticle states. It was noticed by Pushp et al.\cite{pushp} that the
slope of this quantity, in the $\omega \to 0 $ limit,
 while increasing with reduced value of $T_c$
at overdoping, saturates and perhaps even decreases slightly with decreasing
$x$ in the highly underdoped regime. But the low $\omega$
dependence of the DOS determines the low temperature behavior of the
specific heat. For the specific case of Bi2212, we found good
agreement between STS and specific heat data further confirming our work and
providing a strong test of the consistency between experimental data
obtained by these two very different techniques and their consistency with the
resonating valence bond spin liquid.

\begin{acknowledgments}
We have benefitted from
discussions with James LeBlanc.
This work has been supported by the Natural Sciences and Engineering Research
Council of Canada (NSERC)
and by the Canadian Institute for Advanced Research (CIFAR).
\end{acknowledgments}

\end{document}